# Acoustic Metamaterial Beam Splitter

**Liuxian Zhao**

Institute of Sound and Vibration Research, Hefei University of Technology, 193 Tunxi Road, Hefei 230009, China

*Author to whom correspondence should be addressed: lzhao2@nd.edu

**ABSTRACT**

This paper proposes an acoustic beam splitter based on acoustic Luneburg lens. A two-dimensional acoustic Luneburg lens with gradient refractive index is achieved to provide a practical realization for the acoustic beam splitter. The acoustic wave propagation through the beam splitter was explored theoretically, numerically and experimentally. Results show that three output channels can be achieved using a point source on the edge of the beam splitter in a broadband frequency range. Furthermore, the output channel directions can be tuned via changing the design parameter.

# 1. Introduction

Acoustic beam splitter is an acoustic device which is capable of splitting a given acoustic wave with any incidence angle into desired directions, which has been widespread researched with protentional applications in many fields ranging from biomedical imaging to multi-beam acoustic sonar system [1, 2]. In recent years, acoustic beam splitter with improved performance based on acoustic metamaterials have attracted increasing concerns compared with conventional acoustic beam splitter based on natural materials [3, 4]. For example, Graciá-Salgado *et al.* [5] designed a metamaterial with negative density or density near zero to achieve power splitting, the main drawback of the design is the strong transmission losses at the frequencies of negative material properties. Yan *et al.* [6] proposed a zero-index metamaterial to achieve acoustic beam splitter, which can achieve perfect transmission at the output channels. Tang *et al.* [7] proposed an one-way acoustic beam splitter which can divide the input acoustic beam into multiple beams and effectively attenuating the backward reflection. However, the existing acoustic beam splitter designs based on acoustic metamaterial can only work in the very narrow frequency range or at a specific frequency. In addition, the current designs are not omnidirectional, which can only work at specific output directions.

In order to achieve an omnidirectional acoustic beam splitter working in a broadband frequency range, we propose a gradient-index (GRIN) acoustic metamaterial based on acoustic Luneburg lens. The Luneburg lens was first proposed in 1944 [8] in the field of optical wave [9, 10], whose refractive index is spherically symmetric and is able to focus a plane wave at the opposite surface [11, 12]. Jacek Sochacki modified the original Luneburg lens to achieve triple-foci for optical wave [13]. As well-known that the concept of optical Luneburg lens can also been used for the elastic and acoustic fields [14-20]. For elastic wave field, Zhao *et al.* proposed a structural lens based on the optical Luneburg lens, which can split a point source into three collimated waves on a thin plate structure[21]. Similarly, the optical Luneburg lens is applied to the acoustic field for acoustic wave beam splitter in this study, as shown in Figure 1. A point source is used as an incident wave, which is applied at the outer surface of the lens (as indicated **I**). When the acoustic wave interacts with the lens, which can generate three beam channels (as indicated **O₁**, **O₂**, and **O₃**). The distribution of the refractive index of the device is obtained via changing the dimension of unit cell. Our lens has the following attributes: i) the refractive index of the unit cell is independent of the frequency in a broadband frequency range, the lens is broadband, ii) the lens has spherically symmetric gradient index, which has the characteristic of omni-directivity, iii) beam direction can be tuned via changing the design parameter.

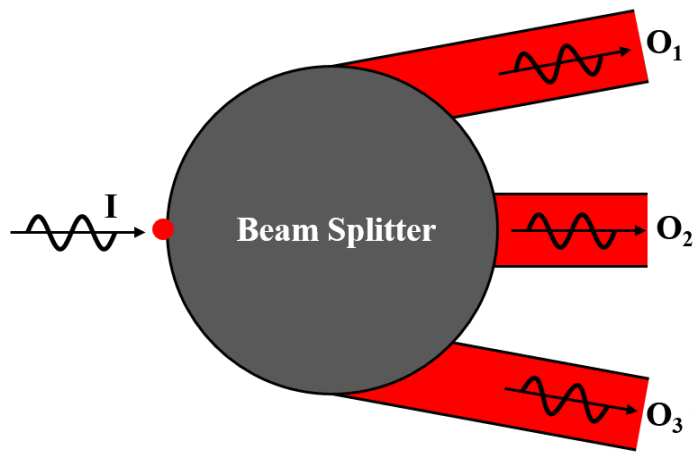

**Figure 1: Schematic of the beam splitter.**

**2. Acoustic Beam Splitter Design**

The acoustic beam splitter is based on the Luneburg lens, which is a GRIN lens with refractive index smoothly varying from the centre of the lens to the outer surface. Considering the beam splitter has two concentric circles with corresponding radius $a$ and $R$. When a line source is used to generate a plane wave, the plane wave passes through the beam splitter can produce triple focusing with focal length $F$ (as indicated in **P₁**, **P₂**, and **P₃**) with distance between neighbouring points $d$, as shown in Figure 2 (a). Inversely, when a point source is placed at the focal length for excitation (such as point **P₂**), three beams can be generated on the opposite side of point source with corresponding beam angles of (-$\theta$, 0°, +$\theta$). The beam splitting angle $\theta$ is a function of parameters $d$ and $F$, which can be expressed as:

$$\theta = \arcsin(\frac{d}{F}) \qquad (1)$$

The corresponding refractive index of the beam splitter is a function of the radial distance $r$ and is expressed as [13]:

$$\begin{cases} n = e^{\left\{\omega(\rho,F) - \frac{\arcsin(\frac{d}{F})}{\pi} \ln \frac{1+\sqrt{1-\rho^2}}{P_a + \sqrt{P_a^2 - \rho^2}}\right\}}, & 0 \leq \rho < P_a \\ n = e^{\left\{\omega(\rho,F) - \frac{\arcsin(\frac{d}{F})}{\pi} \ln \frac{1+\sqrt{1-\rho^2}}{\rho}\right\}}, & P_a \leq \rho \leq 1 \end{cases}, \qquad (2)$$

where $\rho = rn/R$, $P_a$ is a parameter related to the diameter of the inner circle which is between 0 and $R$, $d$ is the vertical distance between the neighbouring focusing points, the range of $d$ is $0 \leq d \leq d_{max}$, here $d_{max} = F\sin\{\pi\omega(P_a, F)/\ln\frac{1+\sqrt{1-P_a^2}}{P_a}\}$. In addition, the parameter $\omega(\rho, F)$ satisfies the following equations [13]:

$$\begin{cases} \omega(\rho, F) = \frac{1}{\pi} \int_0^\rho \frac{\arcsin(k/F)}{\sqrt{k^2 - \rho^2}} dk, & 0 \leq \rho < P_a \\ \omega(\rho, F) = \frac{1}{\pi} \int_\rho^1 \frac{\arcsin(k/F)}{\sqrt{k^2 - \rho^2}} dk, & P_a \leq \rho < 1 \end{cases}, \qquad (3)$$

In Figure 2(b), we give the distributions of $\theta$ with the relation of $d$ at $F = R = 0.1$ m and $P_a = 0.05$. For example, when $d = 0$ m, the splitting beam angle $\theta = 0°$, which indicates that there is only one beam propagates at 0°; and when $d = P_a = 0.05$ m, the splitting beam angle is $\theta = 30°$; note that when $d = d_{max} = 0.0677$ m, which corresponds to the maximum splitting angle of $\theta_{max} = 42.6°$. In addition, the refractive index distributions of the beam splitter with the relation of radial distance $r$ ($d = 0$ m, $d = P_a$, and $d = d_{max}$) are provided in Figure 2(c).

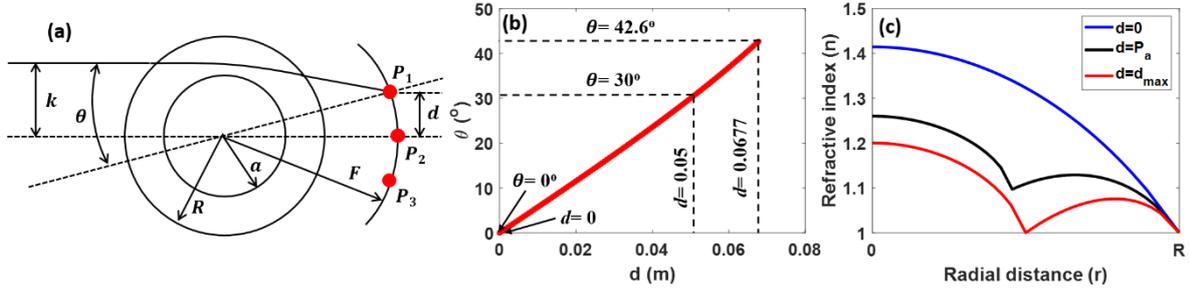

**Figure 2: Mechanism of the beam splitter and its design principles. (a) Schematic of the beam splitter. (b) The relationship between parameter *d* and beam angle *θ*. (c) The refractive index profiles of the beam splitter with different values of *d*.**

### 3. Impedance Matching Beam Splitter

For a proof of concept, an impedance matching beam splitter is used for study. The beam splitter is modelled with continuous change of refractive index based on equation (2). Finite element method was applied to explore the performance of acoustic wave beam splitting. In this study, $F = R = 0.1$ m, $P_a = 0.05$, and different values of $d$ ($d = 0$ m, $d = P_a$, and $d = d_{max}$) were set. In the numerical model, the area of air is $3R \times 4R$. Harmonic analysis was performed using Comsol software at the frequency $f = 15$ kHz. In addition, perfectly matched layers were used to remove the boundary reflection. Simulations were performed with results are provided in Figure 3. It can clearly see that there is only one output beam channel when $d=0$, as shown in Figure 3(a), and the point source can generate three beams when $d>0$, as shown in Figure 3(b)-(c). In addition, the corresponding beam angles are $\theta = 30°$ and $\theta_{max} = 43°$, which are consistent with the theoretical calculations, as shown in Figure 2(b) for $d=0.05$ m and $d = d_{max}$.

Furthermore, the ray tracing method [22] was used to study beam splitting performance based on the Luneburg lens. A bundle of incident rays from a point source on the most left surface of the lens was used for excitation. The calculated ray trajectories were overlapped with the numerical results shown in Figure 3 (a)-(c) for different values of $d$ ($d = 0$ m, $d = P_a$, and $d = d_{max}$). It clearly shows that the ray trajectories agree well with the numerical results.

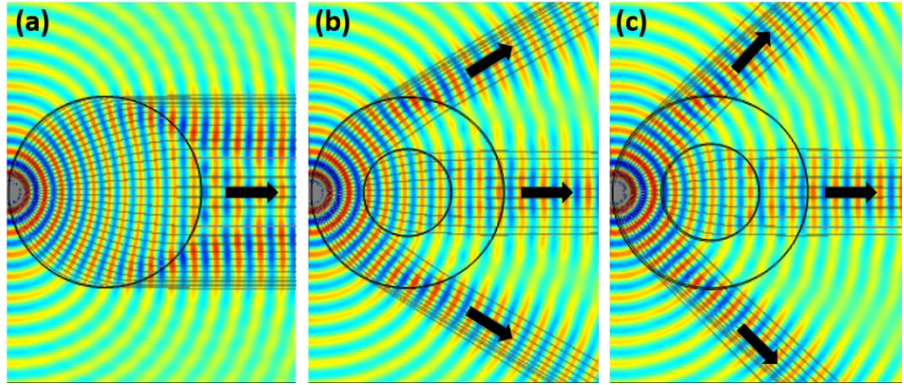

**Figure 3: Numerical simulations and ray trajectories of the impedance-matching lens for beam splitting with different values of $d$. (a)-(c) $d = 0$ m, $d = P_a$, and $d = d_{max}$, respectively.**

**4. Acoustic Metamaterial Beam Splitter**

Impedance matching device is an ideal lens with continuous changing of refractive index, which is impossible to be fabricated with additive manufacturing technique. In order to design a practical acoustic beam splitter, GRIN acoustic metamaterial [23, 24] with graded refractive index was investigated. In this study, an acoustic metamaterial beam splitter was designed with parameters $F = R = 0.1$ m, and $d = P_a = 0.05$ m with the refractive index distribution indicated with black line in Figure 2(c), which is in the range of 1 and 1.26. In this research, 3D truss unit cell was applied to obtain the range of refractive index [25], as shown in Figure 4(a), which is composed of three orthogonal beams. In this study, $D = 5$ mm, the refractive index is a function of the parameter $a$, which is obtained based on the slope of dispersion curve [26], as an example in Figure 4(b). It can be seen from Figure 4(b) that there is almost no change of the slope in the frequency range of 0 and 30 kHz, which means that the refractive index keeps constant in this frequency range. Therefore, the designed acoustic metamaterial beam splitter can work in a broadband frequency range. In this work, we designed a 2D acoustic beam splitter working in the frequency range $f$ = 11 kHz – 17 kHz. In order to reduce the acoustic wave radiation in the thickness direction, five layers in the thickness direction was stacked to form a stable lattice, as shown in Figure 4(c) and (d).

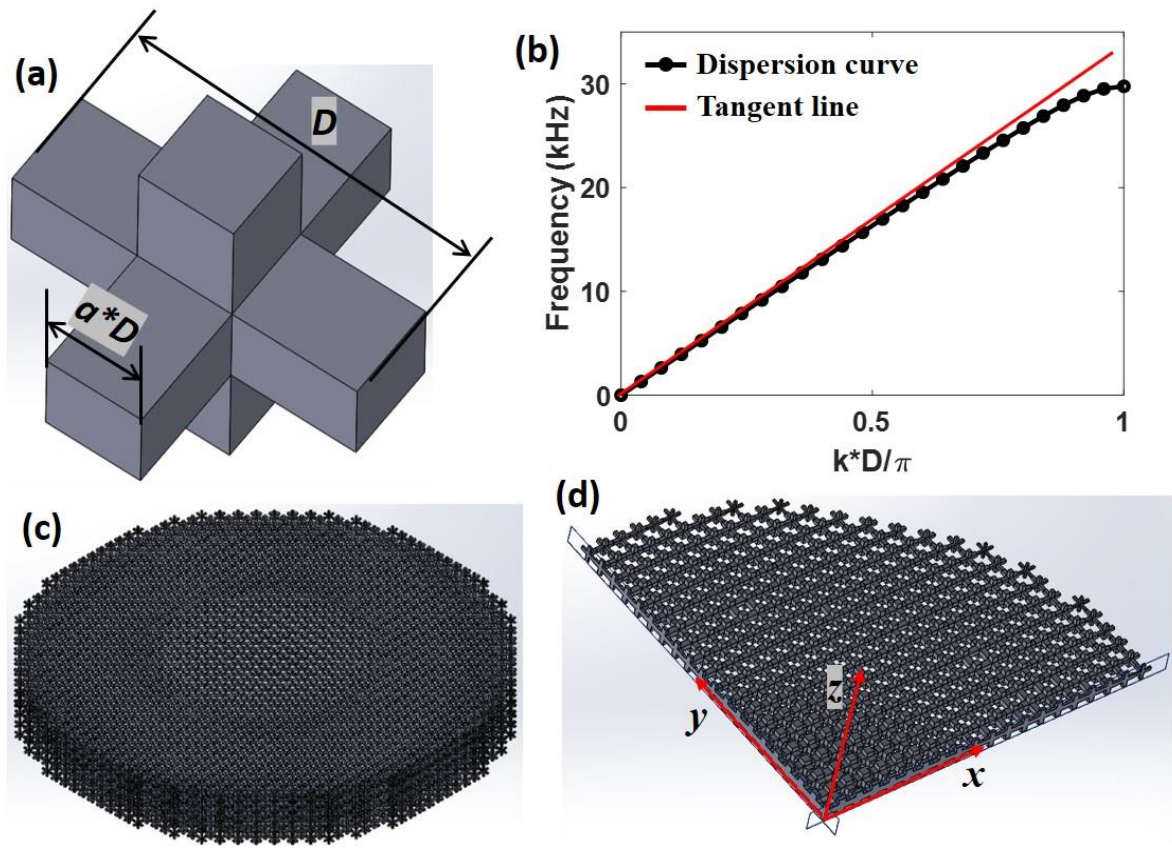

**Figure 4: Acoustic metamaterial beam splitter design. (a) An example of 3D truss unit cell. (b) An example of dispersion curve for the unit cell used to obtain the refractive index. (c) The entire acoustic metamaterial beam splitter with five layers in the thickness direction. (d) Cross-section of the middle layer.**

Equi-frequency contour analysis was performed to analyze the omnidirectional characteristic of the acoustic beam splitter. Different factors of $a$ was analysed to show the anisotropy of the unit cell at the frequency $f$ = 11 kHz, $f$ = 13 kHz, $f$ = 15 kHz, and $f$ = 17 kHz kHz, such as $a$=0.7, as shown in Figure 5. It can be seen that the contour is almost circular for the designed frequencies, which indicates that the splitter is omnidirectional in the working frequencies.

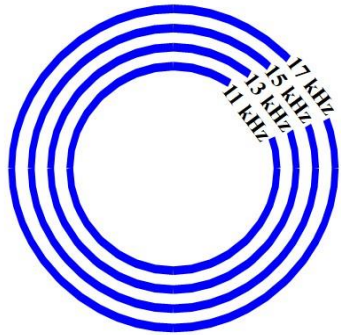

**Figure 5: Equi-frequency contour plot when *a* = 0.7.**

In this research, we explore the performance of the acoustic metamaterial beam splitter. Similarly, harmonic analysis was performed using Comsol software at the frequency $f = 11$ kHz, $f = 13$ kHz, $f = 15$ kHz, and $f = 17$ kHz. In the numerical model, perfectly matched layers were used to remove the boundary reflection. The finite element simulation results are provided in Figure 6(a)-(d) with ray trajectories overlapped, which demonstrates the excellent performance of the beam splitter to convert the point source into three plane waves in a broadband frequency range. In addition, due to the spherical symmetrical design of the acoustic metamaterial, it indicates that the beam splitter can work omnidirectionally. The excitation source was rotated 45º clockwise along *z* axis and all the other simulation settings were kept the same. The simulation results of the rotated excitation source are shown in Figure 6(e)-(h) for the frequency of $f = 11$ kHz, $f = 13$ kHz, $f = 15$ kHz, and $f = 17$ kHz. It can clearly see that the results are similar to the un-rotated model, which verifies the characteristic of omni-directivity. In addition, the splitting angle of the lens is $\theta = 29.5^o$, which is consistent with the theoretical calculations in Figure 2(b) and numerical simulations of perfect impedance-matching lens in Figure 3(b).

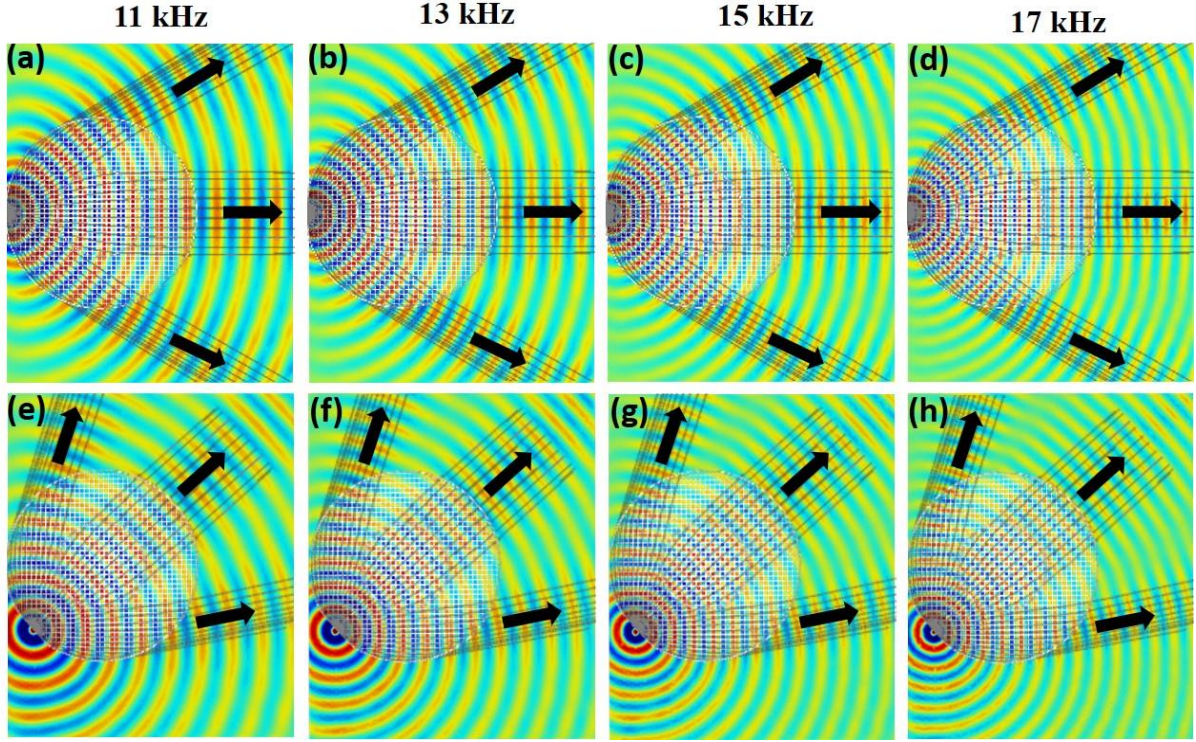

**Figure 6: Numerical simulations of the acoustic metamaterial lens for beam splitting at different frequencies. (a)-(d) Results of the un-rotated excitation source at frequencies $f$ = 11 kHz, $f$ = 13 kHz, $f$ = 15 kHz, and $f$ = 17 kHz, respectively. (e)-(h) Results of the rotated excitation source (45º clockwise along $z$ axis) at frequencies $f$ = 11 kHz, $f$ = 13 kHz, $f$ = 15 kHz, and $f$ = 17 kHz, respectively.**

## 5. Experimental Test of Acoustic Metamaterial Beam Splitter

The acoustic metamaterial beam splitter (Figure 4(c)) was fabricated using 3D printer, as shown in Figure 7(a). The experimental setup used for the device characterization in an anechoic chamber is shown in Figure 7(b). A compact speaker was connected to a tube to generate a point source at the frequency of $f$ = 11 kHz, $f$ = 13 kHz, $f$ = 15 kHz, and $f$ = 17 kHz. A fiber optic sensor was used to collect the sound pressure along the circumferential direction of the lens (the rotation is along z direction).

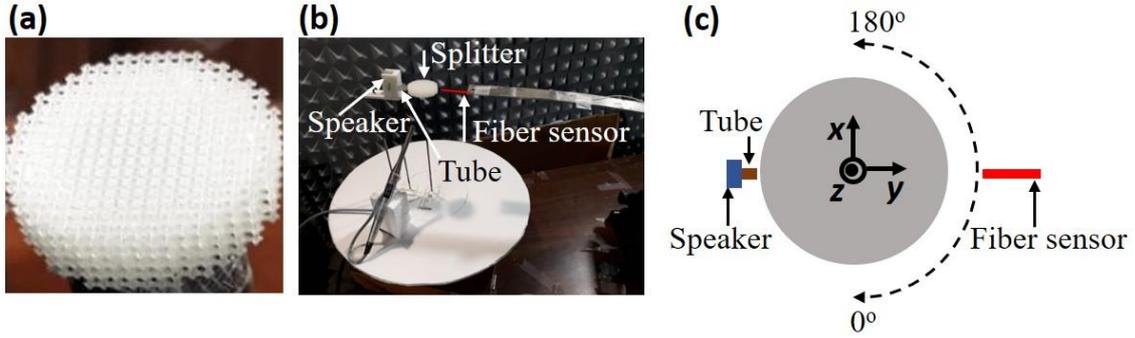

**Figure 7: Experimental setup for characterizing the sound pressure distribution. (a) Photo of the acoustic beam splitter. (b) Photo of the experimental setup. (c) Schematic of the experimental setup.**

We characterized the acoustic beam splitting capability of the acoustic metamaterial beam splitter. A speaker was put at $(x, y, z)=(0, 0, R)$ for excitation, the fiber optical sensor was used to measure the acoustic pressure along the dotted line (Figure 7(c)), and the beam directions were calculated based on the measured acoustic pressure for both un-rotated and rotated excitation sources. The results of the beam directivity are presented in Figure 8. It can be seen that the directivities are around $27°$ for both un-rotated and rotated excitation sources at the frequency of $f = 11$ kHz, $f = 13$ kHz, $f = 15$ kHz, and $f = 17$ kHz, which are consistent with the theoretical and numerical results.

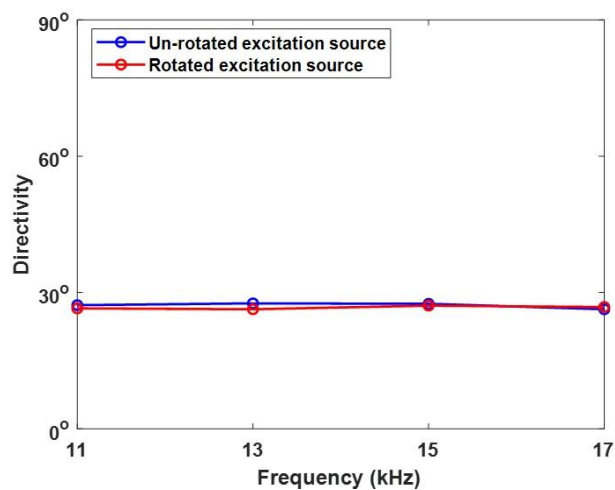

**Figure 8: Experimental result of the acoustic metamaterial lens for beam splitting.**

## 4. Discussion

As presented in the introduction section, the frequency range of the acoustic beam splitter in Graciá-Salgado's research is from 10.1 kHz to 10.4 kHz for the unit cell periodicity $a = 0.005$ m [5], the frequency of acoustic beam splitter in Yan's research is 5.2 kHz [6], and the frequency of acoustic beam splitter in Tang's research is 13.2 kHz for the unit cell periodicity $a = 0.1$ m [7]. In this study, the frequency range of the proposed acoustic beam splitter is from 11 kHz to 17 kHz, which is much wider than the frequency range in the previous studies.

## 5. Conclusions

In conclusion, we show that efficient multiple output channels can be realized by utilizing acoustic beam splitter. Theoretical analysis provides the design principles and obtains the splitting beam angles. Ray trajectories were calculated and compared with the numerical simulations for both impedance-matching beam splitter and acoustic metamaterial beam splitter. In addition, experimental tests were performed and compared with numerical simulations. These results are consistent with each other and verify the beam splittering capability of the proposed splitter. Furthermore, the designed acoustic metamaterial beam splitter works omnidirectional and in a broadband frequency range, which provide the advantage compared with the traditional acoustic beam splitters. We also show that the splitting beam angles can be tuned via changing the design parameter $d$.

**Conflict of Interest**

The authors declare no conflict of interest.